\documentclass[12pt,preprint]{aastex}
\usepackage{amssymb,amsmath,float}

\usepackage{graphicx}

\begin{document}
\title{\large\bf EXTRAORDINARY SOLAR MODULATION EFFECTS
 ON GALACTIC COSMIC RAYS OBSERVED BY V1 NEAR THE HELIOPAUSE }  

\author{W R Webber and $\,^{1}$ and J J Quenby$\,^{2}$}

\affil{\(^{\scriptstyle 1} \){Department of Astronomy, New Mexico
State University, Las Cruces, USA}\\}

\affil{\(^{\scriptstyle 2} \){Blackett Laboratory, Imperial College
London SW7~2BZ, UK}\\}

\begin{abstract}

We discuss here two extraordinary increases of cosmic ray intensity that were
 observed by Voyager1 in the last 1.1 AU before it crossed the heliopause 
in August, 2012, at 121.7AU. These two increases are roughly similar in
 amplitude and result in a total increase of $ \sim1$GV cosmic ray nuclei 
of over 50\% and of 0.01GV electrons of a factor $\sim$ 2. During 
the $1^{st}$ increase the changes
in the magnetic, B field are small. After the $1^{st}$ increase, the B field 
changes become large and during the $2^{nd}$ increae the B field variations and
 the cosmic ray changes are correlated to within $\pm$ one day. The intensity
 variations of H and He nuclei and electrons during these time intervals are
measured from 0.1 to over 1 GV.The total increae of GCR in the two increases
resemble those to be expected from a simple force field "like" solar 
modulation with a modulation potential $ \sim$ 80 MV. This is nearly 1/3 of 
the total modulation potential $\sim$ 250 MV that is required to produce the
modulation of these particles observed at the Earth at the 2009 sunspot minimum
and adds a new aspect to the heliospheric modulation. 
\end{abstract}

\section{INTRODUCTION}

When V1 crossed the heliopause on or about August 25, 2012 (day 238), there were
extraordinary changes in the magnetic field and the energetic particle 
intensities (Burlaga et al., 2013; Stone et al., 2013; Webber and McDonald, 
2013). On that day, the particle intensities and field strength and direction
began a change to values that have remained relatively unchanged now for over
 20 months. Prior to this "final" event there were several unusual features in
 the energetic particle intensities that occured. For energetic particles, we 
mean GCR nuclei and electrons as well as the most energetic anomalous 
cosmic rays
(ACR). The first of these intensity-time features occurred about May $7^{th}$ 
(day 128) when both the GCR nuclei and electrons increased by $\sim$15\% and 
20\% respectively. After reaching these higher levels near the end of May 
(day 150), these intensities remained nearly constant for $\sim$ 58 days 
($\sim$ 2 solar rotation periods). Meanwhile the ACR intensities did the 
opposite, decreasinng by $\sim$20\% to a lower level where they also remained
nearly constant for the 58 day time period.\\
On about July $28^{th}$ (day 210), the GCR nuclei and electron intensities 
increased suddenly for the second time.  This increase was more rapid and
eventually larger (20\% and 40\% respectively) than the $1^{st}$ increase.
However, the increase occured in several stages, the final one starting on 
August $25^{th}$ (day 238). At this time, GCR nuclei increased to their final
values which were $\sim 32$\% higher than they were before May $7^{th}$ for
$>$70 MeV/nuclei  and $\sim$96\% higher for 7-100 MeV electrons. The " trapped"
nuclei, termination shock particles (TSP) and anomalous cosmic rays (ACR),
 disappeared suddenly (Krimigis et al., 2013), so that within just a few weeks
 the intensity of 2 MeV protons was less than 0.1\% of their intensity
 before May $7^{th}$.\\
The magnetic field  changes, both in amplitude and direction, are a crucial
 backdrop for the energetic particle changes. In this paper, we will discuss
 the GCR and magnetic field temporal changes during the time period from day
128 to day 238 when V1 moved outward $\sim$ 1.1 AU. We will then present the
rigidity dependence of the observed changes for the species H, He and electrons
to help define the modulation mechanisms responsible for the intensity changes.

\section{THE INTENSITY-TIME CHANGES AND A DISCUSSION OF THEIR IMPLICATIONS}

The data presented here from V1 clearly shows two periods of increase for both
GCR nuclei and electrons. This is illustated in Figure 1 which shows the
 integral rate of $>$70 MeV nuclei and 5-60 MeV electrons. The total increase
 for each component from before May $7^{th}$ to after August 25$^{th}$
(32\% for $>70$ MeV H and 96\% for 5-60 MeV electrons) are made equal in 
the plot
using different scales on the left hand and right hand axes. This is to show
the relative magnitude of the increase on May $8^{th}$ to that in the second
 change between July $28^{th}$ and August $25^{th}$ for the two species.
The intensity changes for the two species are not identical in relative 
magnitude or in their relationship to the magnetic field.\\

\begin{figure}
\includegraphics[width=\linewidth]{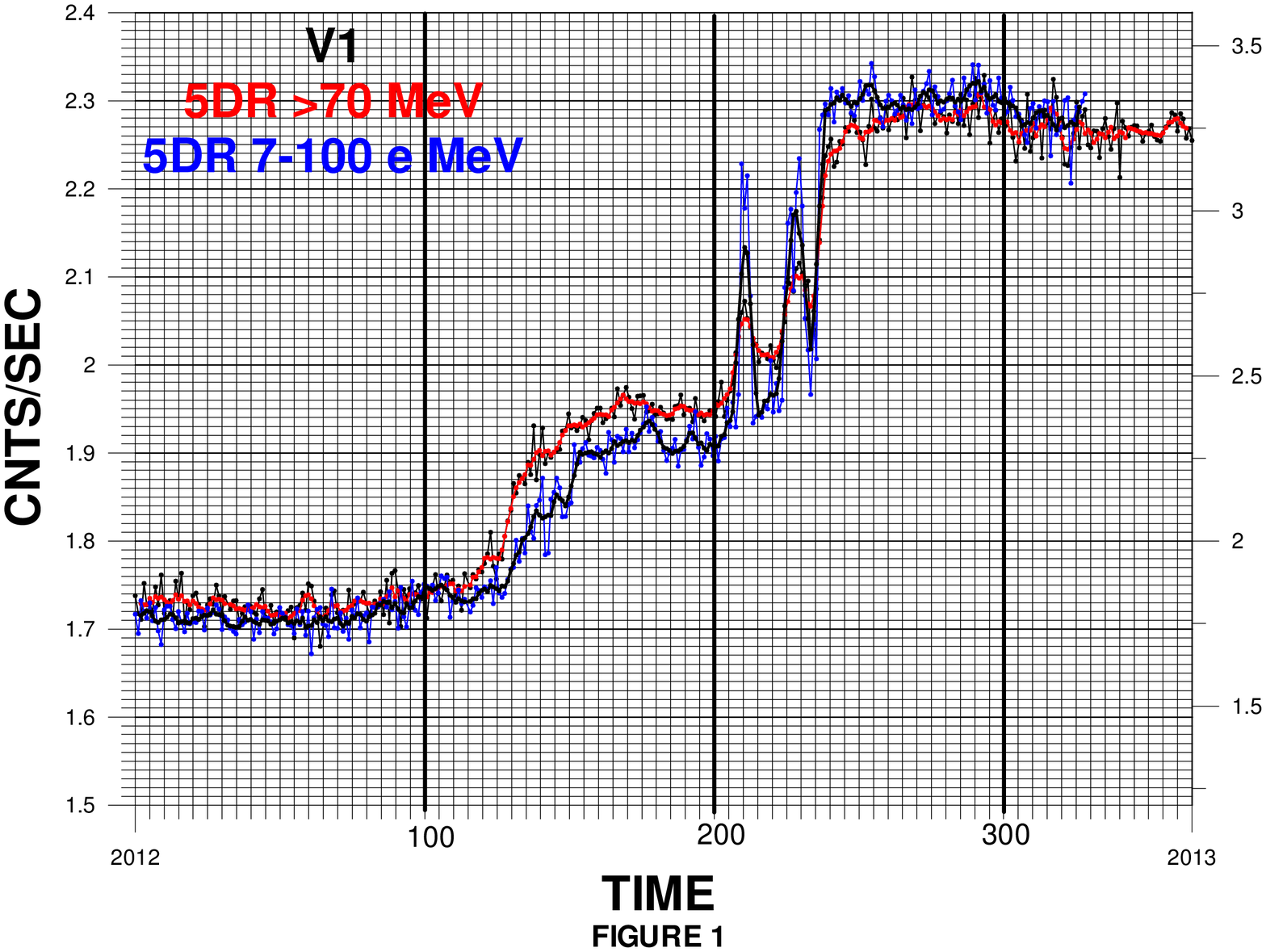}
\caption[GCR Nuclei and Electron Variations Near Boundary]
{Five day GCR running average of $>70$ MeV (mostly nuclei, left scale), red line
 and 5-60 MeV (mostly electrons, right scale), blue line. 
The two increases starting 
day 128 and day 208 are the solar modulation events discussed in the text}
\end{figure}

The format in Figure 2 is similar to Figure 1 and shows the electrons 3-10 Mev 
(left hand scale and the relative magnetic energy density $\sim B^{2}$ (right
 hand scale).

\begin{figure}
\includegraphics[width=\linewidth]{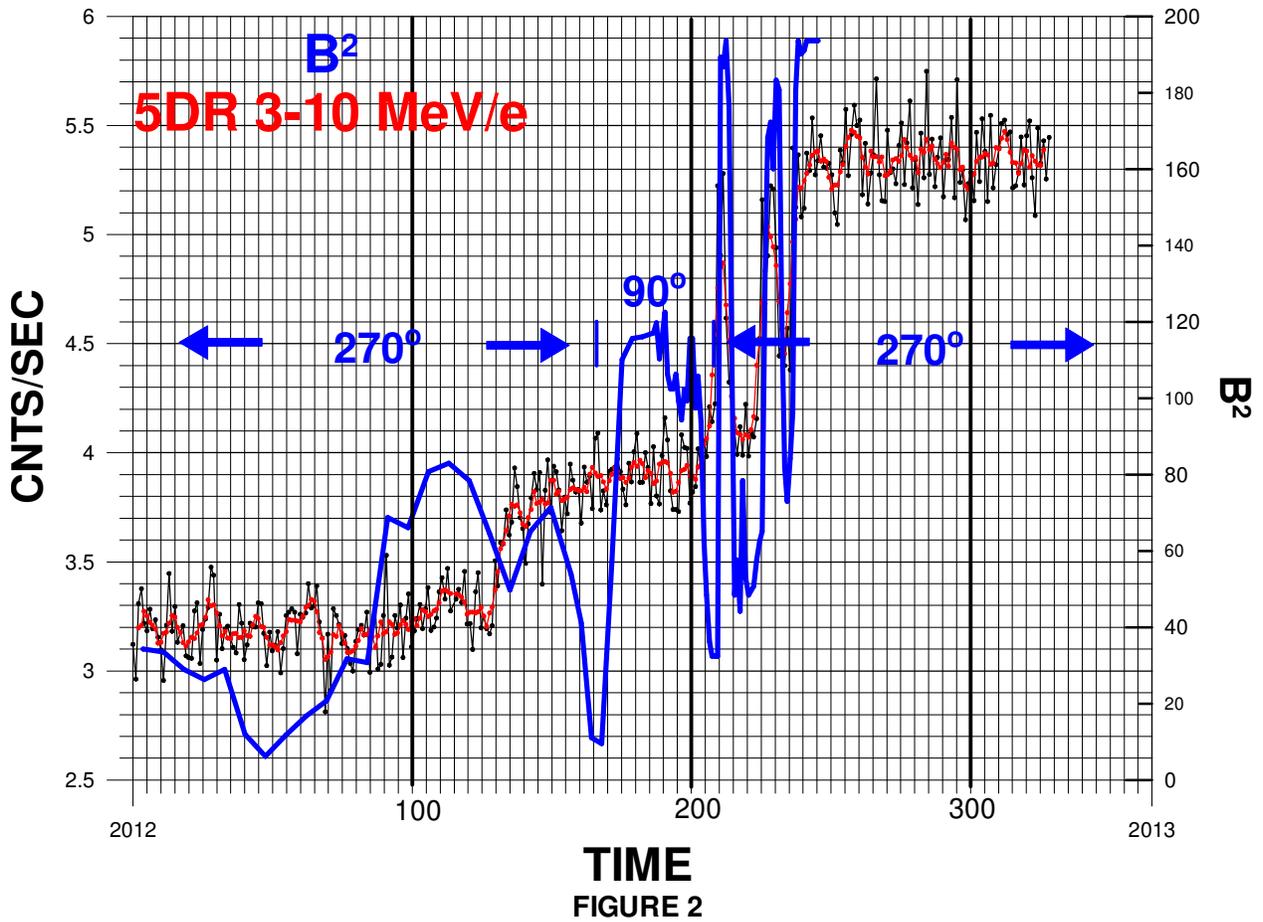}
\caption[GCR Electron and Field Variations Near Boundary]
{Similar to figure 1 except the 3-10 MeV (mostly electrons) GCR are shown
in red, along with the relative energy density ($B^{2}$) and direction of the
magnetic field.}
\end{figure}

 For the 3-10 MeV electrons, the total increase is 69\%, which is
 less than the increase of 96\%  for the 5-60 MeV electrons.\\
For the B field, also shown in Figure 2, the changes in amplitude are over a 
factor 4 during this time period from values $\sim1 \mu $G to $\sim 4.5\mu $G
(the final field value)  in just a few days from day 208-210. There are two 
changes in B field direction (days 163-172) from a positive to a negative
 polarity and a much more sudden and final change, on days 208-209, from
negative to positive polarity.\\
The $1^{st}$ step in the cosmic ray intensity changes at day 128 does not occur
in association with any major B field amplitude change. After this intensity
 change and later during the period of constant GCR intensity, the B amplitude
decreases and then increases by a factor of 3 between days 162-172, during a 10
day period when the field direction is also changing from $270^{\circ}$ to 
$90^{\circ}$. During this 10 day period the field inclination increases from
$\sim 0^{\circ}$ to $\sim 90^{\circ}$ (Burlaga et al., 2013). The locally 
measured GCR intensities were remarkably insensitive to these extraordinary B
field changes.\\
The next field polarity change occured on day 209 and is the final, decisive
polarity change from $90^{\circ}$ to $\sim 270^{\circ}$. 
$\underline{One \: day \: later}$ the B field amplitude changes by a factor 3 in one
 day to its final value of $4.5\mu $G. The B field changes that occur between
days  210 and 238 are matched within $ \pm$ day by the corresponding increases
in GCR and decreases in TSP and ACR as seen in Figure 2 and also in Krimigis
et al., (2013) and Burlaga et al., (2013). Also, as seen in Figure 2, the lower
 rigidity electrons are more responsive than the nuclei to these changes in
field amplitude that pass V1 between day 210 and day 238.\\
So overall we have the observation by V1 that the $1^{st}$ GCR increase
 stating on day 128 was $\underline{not}$ coincident with corresponding large 
B field amplitude or direction changes. A following period of nearly constant
 GCR intensity, however, was coincident with large amplitude and direction
 changes of the B field, as well as unusual field elevation angle changes.\\    The $2^{nd}$ GCR increase starting on day 208 and the following intensity 
changes culminating with the final increase on day 238, were all simultaneous
with $\pm$ one day with the very large B field magnitude changes, but there 
were no field direction changes during this time. The lower rigidity electrons
 were more responsive to these B field changes than the GCR nuclei whose 
rigidity is $\sim 50$ times that of the electrons. In this second increase of
 GCR, the TSP and ACR intensity changes were opposite to the GCR to within 
$\pm$ 1 day as these components disappeared.  

\section{INTENSITY AND SPECTRAL CHANGES OF GCR H, He NUCLEI AND ELECTRONS
 BETWEEN DAY 128 AND DAY 238 OF 2012}

In the Stone et al., (2013) article in Figures 2, 3 and 4 the intensities and 
spectra of H and He nuclei and electrons are shown for the time periods before
May $8^{th}$ and after August $25^{th}$. This includes the period of the two 
GCR increases. Below an energy of $\sim80$ MeV/nuc., the time period before 
May $8^{th}$ is contaminated by background ACR intensities for H and He nuclei
and therefore these energies cannot be used  in the comparison.\\
The intensity changes of these particles with different mass to charge ratio,
A/Z, have historically been very useful for understanding the origin of solar
modulation effects. For example, Gleeson and Axford (1968) have compared the
H and He intensity changes in their derivation of the force field 
approximation to the solar modulation. They find that if the changes in 
intensity, $M=\beta ln (j_{1}(P)/j_{2}(P)$, are
 plotted as a function of rigidity,
 there is a splitting of the modulation for each species according to their
charge to mass ratio A/Z. This splitting arises from the fact that the 
modulation itself, expressed in MeV, is defined by a modulation function,
$\Psi=Ze\int ((V/3)dr/K(r)$ where $K(r)$ equals the scaler diffusion 
coefficient which has dependence $K(r) \propto \beta P f(r)$ and V is the 
radial wind speed. Gleeson and Axford (1968), however, also introduced another
 quantity called the modulation potential, $\phi=\int(V/3)dr/K(r)$ expressed
in MV. This modulation potential is the same for H, He and electrons at the 
same rigidity. The modulation function, M, that is commonly used, is defined
by $\Psi$. M describes the amount of modulation between two different times
(or places) and is different at the same rigidity for particles of different
A/Z. Hence the use of the term charge splitting when this quantity is used.
The section by Ken McCracken pp 50-58 in the book Cosmological Radio Nuclei
(2012) discusses the two quantities, modulation function and modulation 
potential. The modulation function is useful for comparing intensity changes
of different species.\\

\begin{figure}
\includegraphics[width=\linewidth]{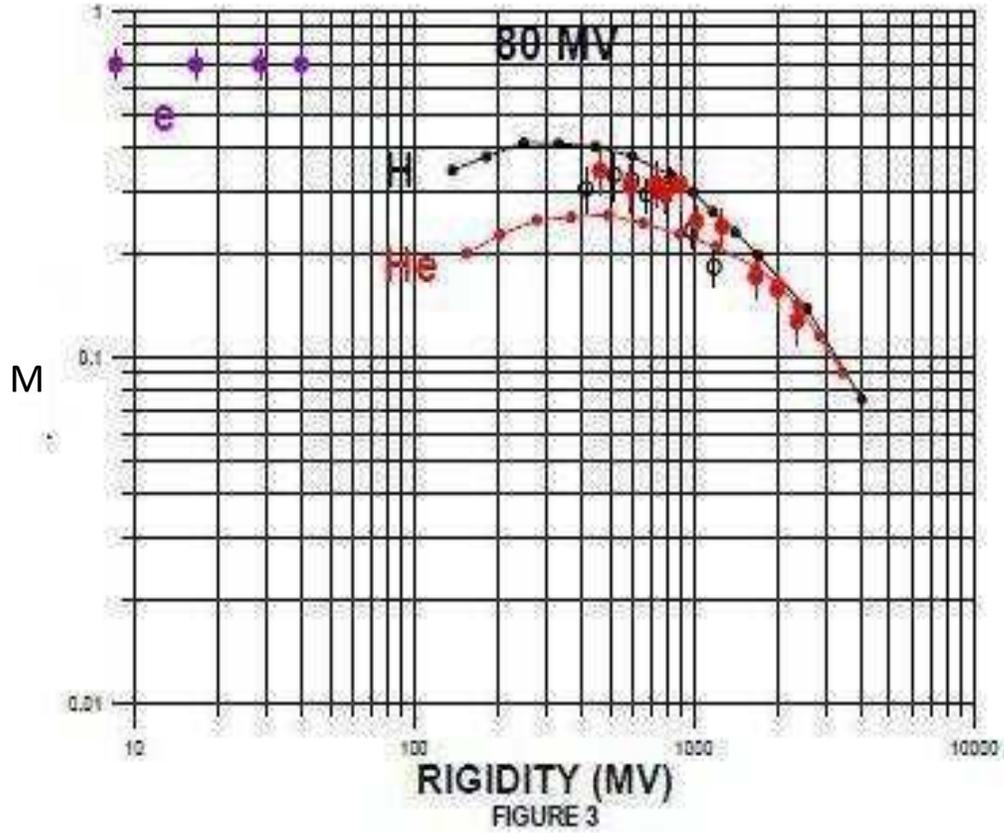}
\caption[Modulation Function] {The modulation function 
$M=\beta ln(j_{1}/j_{2})$ calculated for a 
modulation potential=80 MeV (solid lines) and the observed modulation obtained 
by comparing the H (black), He (red) and electron (blue) spectral intensities
measured at V1 before day 128 and after day 238 of 2012. The details are 
discussed in the paper.}
\end{figure}
Figure 3 is a plot of the expected variation of $\beta ln(j_{2}/j_{1})$
with P for H and He where the solar modulation potential is taken to be 80 MV.
Note the charge splitting of the amount of solar modulation which results in
a 2 times greater modulation for H than He at rigidities $\leq 1$ GV. The data
 points are for the observed modulation of H and He from before May $8^{th}$ to
after August $25^{th}$.They are roughly consistent with an overall
 mododulation $\sim$ 80 MV, but with no obvious charge splitting. The observed 
modulation function for electrons at lower rigidities is independent of P and
 is $\sim2$ times that for H and He nuclei at $\sim0.5$ GV.

\section{GENERAL COMMENTS}

It is not the intent of this paper to develop a theoretical model for an
explanation of these modulation effects observed by V1 by the CRS instrument.
We believe that the GCR intensity changes are so unusual and unprecedented in 
the history of cosmic ray studies that they are not easily accomodated within
the Parker (1963) heliospheric modulation picture as developed by many others,
 e.g. Gleeson and Axford, 1968; Fisk and Axford, 1969. But there are certain 
features of the modulation that may indicate the characteristics of the B field,
plasma flows, etc., that affect the entrance of the Local Interstellar Spectrum 
(LIS) cosmic rays into the heliosphere.\\
We recognise that the $1^{st}$ step in this modulation process is related to
 the $1^{st}$ event that  started about May $8^{th}$ (DOY 128). This event 
contributed about 40\% of the total increase for both GCR nuclei at higher 
rigidties and electrons at lower rigidities. The changes in the B field were
small during the 10-20 day time period of the $1^{st}$ increase as noted
 earlier. During the following $\sim58$ day time period up to about day 200
the intensity of the GCR remained relatively constant to within a few percent.
However the B field recorded some of the largest changes yet seen at V1 along
with a polarity from $270^{\circ}\rightarrow90^{\circ}$ and along with
unprecedented changes in the elevation angle near the end of this constant
GCR intensity period.\\
The lack of significant time correlation between the GCR intensity changes and
 the B field changes during the entire time period from DOY 128-208 suggests
 that the GCR changes during the first increase could be related to much larger
scale features that may not be evident in the local field being measured at
V1 at that time.\\
The lack of correlation between B and the GCR intensities in the $1^{st}$
event is definitely not present in the $2^{nd}$ event which started on July 
$28^{th}$ (day 208), In this event, from July $28^{th}$ to August $25^{th}$,
the GCR and B field changes were correlated to within $\pm \sim 1$ day or less.
This correlation continues through all 5 intensity increases and decreases, all
of them exceptional, until the final increase on August $28^{th}$ (day 238). In
each of the increases the B field magnitude and direction at the times of the 
B field maxima was essentially the same as that observed after August $28^{th}$.
The GCR electrons and nuclei, however,  did not reach intensities that were
observed after August $28^{th}$. For electrons from 5-60 MeV the peak increases
were $\sim80$\% of the post August $28^{th}$ intensity. For nuclei, the 
increases reached levels $\sim50-60$\% of the post August $28^{th}$ intensity.
So there is a distinct rigidity dependence of the GCR distribution within these
structures that pass V1. Note that the speed of V1 is $\sim0.01$ AU/day. So the
$\pm$ day correlation between B and GCR could have a scale $\sim0.01$ AU!\\
The intensity changes of electrons, H and He nuclei as a functiom of P for the
overall time period from day 128 to day 238  are large and well defined. At 
least two features of the $\beta .ln (j_{LIS}/j_{2})$ vs P data for this
modulation of the different species, shown in Figure 3, are important:
(1) The intensity changes of electrons are nearly independent of rigidity at
low rigidities. In addition, if the values of electron modulation function
$\beta .ln(j_{LIS}/j_{2})$ at low rigidities is extrapolated to higher 
rigidities, it has a value $\sim2$ times the value of the modulation function
observed for H and He at about 0.5 GV. (2) The H and He nuclei have observed
 values of the modulation function that are similar. This lack of splitting in
 the modulation of these different charges is not apparent in the normal 
heliospheric modulation for the Gleeson and Axford (1968), force field
 approximation (eg Lezniak and Webber, 1971).\\
In a simple spherical modu;ation model (eg., Gleeson and Axford, 1968) the 
average magnitude of the modulation function for H and He as a function of
 rigidity for this final time interval would be equivalent to that for a 
modulation potential equal to 80 MV, as indicated in figure 3. In fact, in this
simple picture, a modulation potential equal to 250 MV, starting with the LIS
spectra observed by V1, will reproduce the Carbon and heavier nuclei spectra
measured by ACE in 2009-2010 at the Earth (eg., Lave et al., 2013) as well as
 the spectrum observed by PAMELA in 2009 (Potgieter et al., 2014). Thus the
modulation observed by Voyager 1 in the last 1.1 AU of the heliosheath is an
important contributor to the overall solar modulation observed at the Earth,
contributing as much as $\sim 1/3$ of the modulation potential observed at the
Earth at this time of the solar cycle.

\section{SUMMARY AND CONCLUSIONS}

This paper describes two large and unprecedented modulation events of GCR 
observed at V1 starting on May $8^{th}$ and July $28^{th}$ just prior to the
crossing of the heliopause on August 25, 2012 at a distance of 121.7 AU. These
events resulted in the increase of GCR electrons from 5-60 MeV by a factor 
$\sim2$ and H and He nuclei above $\sim0.5$ GeV to increase by factors up to 3
at the lowest rigidities. Although these increases are complex in temporal
stucture, they can be represented by a modulation potential change $\sim 80$ MV
which is 1/3 of the total solar modulation required to reproduce the spectra of
 the same nuclei observed at Earth at a time of sunspot minimum in 2009
(eg., Mewaldt et al., 2010). Thus a new and significant feature is added to the
 description of solar modulation in the heliosphere.\\
The first modulation event occured when V1 was 1.1 AU inside the HP. The
 intensity increases starting on May $8^{th}$ (day 128) and continuing up to
day 150 amounted to about 40\% of the combined increase for both events. During
this GCR increase there were only modest changes, both negative and positive, 
in the B field amplitiude with no change in direction. 
For the next 60 days 
the GCR and ACR intensities remained almost constant. However, between days
150-160 the B field changed direction from $270^{\circ}$ to $ 90^{\circ}$  and
then decreased by a factor $\sim 2.0 $, followed in a few days with a sudden 
increase by a factor $\sim3$ accompanied by an increase in elevation angle from
$\sim 0^{\circ}$ to $90^{\circ}$. It almost seems like the B field was 'turning
its self inside out' in a period of a few days, perhaps due to 
the passge of a very large
scale quasi-periodic structure, but without any observable effects on GCR or 
ACR.\\
The time period of $90^{\circ}$ polarity ended suddenly on July $28^{th}$ (day
208) when the polarity changed to $270^{\circ}$ followed by an increase in the 
magnitude ot the B field (on day 209), again by a factor $\sim 3$ to essentially
 its final value $\sim4.5 \mu G$ after day 238. This increase on July $ 28^{th}$
and the subsequent changes in B were coincident within 1 day with corresponding
 positive and negaive intensity changes of GCR. In this period the changes in 
ACR (Stone et al., 2013) and TSP (Krimigis et al., 2013 ) were exactly opposite
to those of GCR. The details of these changes provide a glimpse into features
of the heliopause with structures with a scale which could be $\sim0.01$AU.
These structures could be moving with speeds much greater than V1.\\
The second modulation increase was embedded in massive B field fluctuations 
in contrast
to the first modulation increase in which these B field changes were small. In
the first increase the B field magnitude and polarity changes actually occurred
$\underline{after}$ the increase and when the GCR changes themselves were
 small. The $1^{st}$ and $2^{nd}$ GCR increases have roughly the same magnitude
and same rigidity dependence, however, despite their greatly different 
correlation with the B field. They could be part of a larger structure $\sim1$
AU in extent that characterizes the heliopause region. An over-riding feature
of this data is the complexity of the changes of both the B field and the GCR
 nuclei and electrons on scales which could be as small as $\sim 0.01$AU. The
correlation and lack of correlation of these changes is evident in the second
 and first modulation events, respectively. A companion work (Quenby and Webber,
2014) attempts to provide a model to explain the events just described, 
concentrating on estimates of likely plasma speed and diffusion parameters
thought to be necessary to account for the V1 GCR observations.\\
In a astrophysical sense our heliopause, which may be typical of 
millions of similar cases in the galaxy where there are stellar winds existing
on all kinds of scales, is notable for the energy it removes from the local 
GCR rather than the acceleration of any particular particle population.\\

\section{ACKNOWLEDGEMENTS}

W.R. Webber wishes to thank his Voyager colleagues from the CRS instrument,
 Project PI Ed Stone, Alan Cummings, Nand Lal, Bryant Heikkla and the late 
Franck McDonald. Support from JPL is greatly appreciated.

\section{REFERENCES}
\noindent
Beer J., McCracken K., von Steiger R., 2013, Cosmogenic\\ 
\indent Nuclides. Springer, Berlin\\
\noindent
Burlaga L. F., et al., 2013, Science, 340, 150\\
\noindent
Fisk L. A., Axford W. I., 1969, J. Geophys. Res, 74, 4973\\
\noindent
Gleeson L. J., Axford W. I., 1968, ApJ, 154, 1012\\
\noindent
Krimigis S. M., et al., 2013, Science 344, 144\\
Lave K. A., et al., 2013, ApJ, 770, 117\\
\noindent
Lezniak J. A., Webber W. R., 1971, J. Geophys Res.,\\ 
\indent
76, 1805\\
\noindent
Potgieter M S., et al., 2014 COSPAR Moscow\\
\noindent
Quenby J J., Webber W R., 2014, submitted for publication\\
\noindent
Stone E. C., et al., Science, 341, 150\\
\noindent
Parker E W., 1958, Phys. Rev., 110, 1445\\
\noindent
Webber W R., McDonald F B., 2013, Greophys. Res. Lett.,\\
\indent
 40, 1665\\
             
\end{document}